\documentclass[twocolumn,superscriptaddress,aps]{revtex4}

\usepackage{graphicx}
\usepackage{dcolumn}
\usepackage{bm}

\begin{document}


\title{An experimental setup for high resolution 10.5~eV laser-based angle-resolved photoelectron spectroscopy using a time-of-flight electron analyzer}

\author{M.~H.~Berntsen}
\affiliation{Materials Physics, KTH Royal Institute of Technology, S-16440 Kista, Sweden}
\author{O.~G\"{o}tberg}
\affiliation{Materials Physics, KTH Royal Institute of Technology, S-16440 Kista, Sweden}
\author{O.~Tjernberg}
\email{oscar@kth.se}
\affiliation{Materials Physics, KTH Royal Institute of Technology, S-16440 Kista, Sweden}

\date{\today}

\begin{abstract}
We present an experimental setup for laser-based angle-resolved time-of-flight (LARTOF) photoemission. Using a picosecond pulsed laser, photons of energy 10.5~eV are generated through higher harmonic generation in xenon. The high repetition rate of the light source, variable between 0.2-8~MHz, enables high photoelectron count rates and short acquisition times. By using a Time-of-Flight (ToF) analyzer with angle-resolving capabilities electrons emitted from the sample within a circular cone of up to $\pm15$ degrees can be collected. Hence, simultaneous acquisition of photoemission data for a complete area of the Brillouin zone is possible. The current photon energy enables bulk sensitive measurements, high angular resolution and the resulting covered momentum space is large enough to enclose the entire Brillouin zone in cuprate high-$T_{c}$ superconductors. Fermi edge measurements on polycrystalline Au shows an energy resolution better than 5~meV. Data from a test measurement of the Au(111) surface state is presented along with measurements of the Fermi surface of the high-$T_{c}$ superconductor Bi$_{2}$Sr$_{2}$CaCu$_{2}$O$_{8+\delta}$ (Bi2212).
\end{abstract}

\pacs{}
\maketitle

\section{Introduction}

During the last decades angle resolved photoelectron spectroscopy (ARPES) has developed into one of the leading experimental techniques for studying the electronic structure in low-dimensional systems \cite{Damascelli03}. In particular, the development of third generation synchrotron radiation sources and the advent of hemispherical electron energy analyzers with two-dimensional detection capabilities have paved the way for ever increasing energy and momentum resolution, thus giving access to the finer details in the electronic structure in complex materials. At present, some of the systems heavily studied by the ARPES community are the high-temperature superconductors (HTSC) in which, in some cases, the superconducting gap is of milli-electronvolt (meV) size, thus motivating the need for instruments with high resolution in both energy and momentum. Recent advances in laser based light sources have enabled a number of groups to realize new, state of the art, laboratory based ARPES setups \cite{Koralek07,Liu08,Kiss08}. Due to the narrow linewidth and low photon energy of the lasers, combined with spot sizes below 200~$\mu$m, these setups achieve phenomenal resolution in momentum and energy. Although very competitive and highly valuable for further studies of the electronic structure in exotic solids, these high-resolution systems have a few limitations. Firstly, the low incident photon energy ($<$~7 eV) enables only a part of the Brillouin zone (BZ) in many materials of strong current interest, e.g. HTSC, to be probed. In order to cover the entire BZ higher photon energies are needed. Secondly, the hemispherical analyzers measure along a line in momentum space ($E(k_{x})$) which means that mapping of areas of the BZ involves rotating the sample relative the analyzer to cover the emission angles of interest. The angular, and consequently the momentum, resolution perpendicular to the axis of rotation is limited by the analyzer entrance slit and the precision with which the rotation is performed. This mapping procedure is a tedious task and can be a limiting factor when studying, e.g. spectral weight transfer, or systems where one would like to follow the band dispersion with high resolution not only in the traditional energy-momentum picture but also directly observe the  momentum-momentum band dispersion.

In this article, we present an ARPES system which addresses these two issues. The current system uses a laser based light source which through higher-harmonic-generation (HHG) produces photons with energy 10.5~eV. This provides a substantial increase in kinetic energy of the photoelectrons compared to the $<$~7 eV systems, accompanied by an increase in covered momentum space. Thus, for the cuprate HTSCs the complete Brillouin zone is covered. In addition, the electron analyzer is of the Time-of-Flight (ToF) type which provides three dimensional detection capabilities. This gives the possibility of collecting energy-momentum data for a complete area $E(k_{x},k_{y})$ of the Brillouin zone rather than along a line $E(k_{x})$, as with hemispherical analyzers. It is believed that these developments will provide completely new experimental possibilities and insight to the nature of complex solids. 


\section{Laser based ARPES using Time-of-Flight analyzer}

The experimental setup described here was designed based on the following criteria; The ARPES system should: i) be able to collect photoemission data for complete areas of the Brillouin zone, ii) have the highest possible resolution in angle and energy, iii) reach the Brillouin zone edge in most materials of strong current interest. In order to simultaneously acquire data from an area of the BZ an angle-resolving time-of-flight (ARTOF) electron analyzer is required. The ARTOF is not simply a traditional drift tube but rather an electrostatic lens (with some defined angular acceptance) which images the emitted electrons onto a detector positioned at the end of the lens. The emission angle from the sample and kinetic energy of an electron is given by a combination of its spatial striking position on the detector and flight time. Since the ARTOF measures the flight times of electrons the system requires a pulsed light source. To achieve high angular resolution, low incident photon energies are desirable since the angular spread of spectral features will be wide at low kinetic energies. In order to ensure high energy resolution the spectral linewidth of the light source has to be narrow and the repetition rate of the source sufficiently high so that reasonable photoelectron count rates can be achieved without running into space charge effects as a result of too many photoelectrons being generated per pulse \cite{Zhou05,Passlack06}. Reaching the BZ edge in, for example, HTSCs requires photon energies around 10~eV. However, too high photon energies reduces the angular resolution, therefore a compromise between available momentum space and angular resolution has to be made. Considering the above mentioned criteria a natural choice is to use a laser as light source since it provides short pulses with a well defined energy and polarization which simplifies the timing issue and enables high energy resolution. 

Lasers have been used for some time as light sources in photoemission \cite{Haight94,Mathias07}, mostly for time-resolved experiments \cite{Haight88, Karlsson96, Carpene09} (pump-probe) using drift tube time-of-flight (ToF) analyzers with single or multiple detector anodes, resulting in systems with energy resolution of several 10s of meV at the best. Higher-harmonic-generation (HHG) in noble gases has enabled photons in a wide range of energies from 10-80 eV \cite{Haight94}. Since these sources traditionally have been used for pump-probe setups the pulse length has been short (typically subpicosecond,except ref.~7) to ensure good time-resolution. The short pulses have provided the high pulse energies required for HHG but the repetition rates of the sources have been very low, usually $\leq$~10~kHz \cite{Dakovski10}. For an ARTOF, as the one described in this article, these low repetition rates would lead to very long acquisition times and limited angular and energy resolution due to the low photoelectron intensity. Hence, these traditional light sources are not suitable for a high resolution ARTOF setup. One way to increase the count rate is of course to increase the number of photons (and hence electrons) generated per pulse. This is, however, not desirable since one quickly runs into space charge effects \cite{Passlack06}. Ideally one would like to generate few electrons per pulse and instead use as high repetition rate as possible. On the other hand, higher repetition rates means that the pulse energy is reduced and it is only recently that lasers with repetition rates greater than a few 10s of kHz and pulse energies large enough for HHG have been available. Also, the quasi-continuous lasers used by recent high resolution ARPES setups \cite{Koralek07,Liu08,Kiss08} have too high repetition rates (80-100 MHz) for a ToF analyzer but works fine with a hemispherical analyzer since timing of the light source for this type of analyzer is not a necessity.  

The setup described in this article works at a photon energy of 10.5~eV and has a tunable repetition rate in the range 0.2-8 MHz which is at least one order of magnitude higher than previous $>$~7 eV setups. The high pulse energy of the light source in combination with the high repetition rate provides a large photon flux ($>$~$10^{12}$~photons/sec) which enables short acquisition times and at the same time minimizing space charge effects. The present system shows that the limit of high resolution, high repetition rate, laser based ARPES can be pushed to 10.5~eV photon energy. By increasing the photon energy to 10.5~eV one covers the entire BZ of cuprate HTSCs without sacrificing too much of the angular resolution compared to the $<$~7~eV setups. One additional advantage coupled to photoemission experiments at low photon energies, i.e. low kinetic energies of the electrons, is the increased bulk sensitivity. For ARPES measurements, kinetic energies in the range 20-100~eV are frequently used. At these energies the mean free path of electrons is typically $<$~10~\AA \cite{Seah79}, which makes the technique very surface sensitive. For around 10~eV kinetic energy the mean free path is approximately 10-50~\AA, depending on the material. Consequently, with the present system high resolution, bulk sensitive ARPES measurements of, for example, the Fermi surface (FS) and the electronic structure near the Fermi level in HTSCs over a complete area of the Brillouin zone can be performed. We believe that this will provide valuable pieces of information to the studies of high-temperature superconductivity in these materials.


\section{System overview}

\subsection{Light-source and monochromator}

\begin{figure}%
\includegraphics[width=8.5cm]{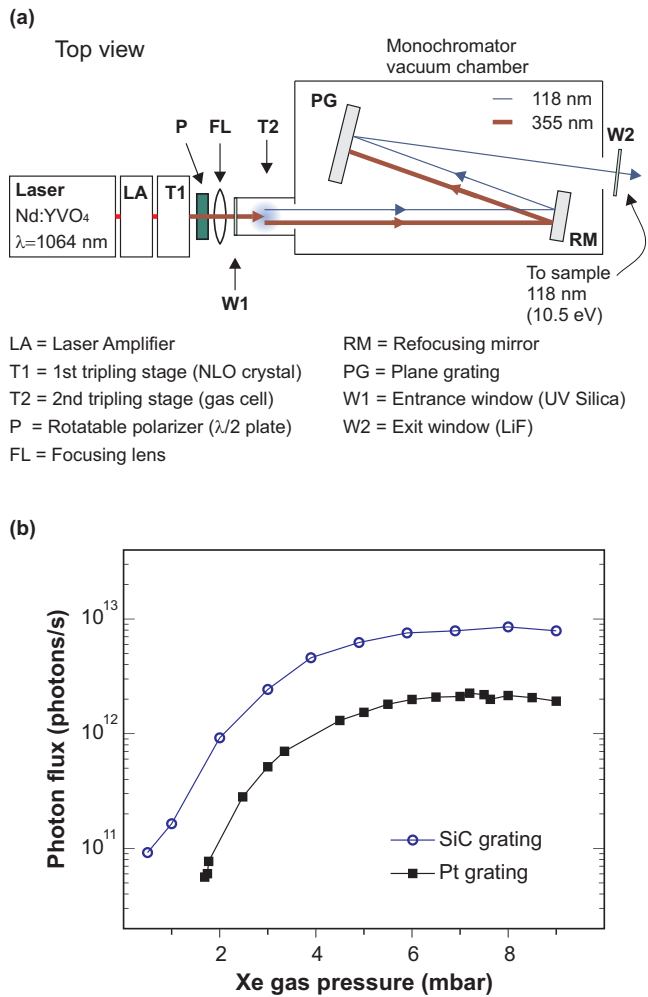}%
\caption{(Color online) (a) Schematic overview of the light source and monochromator showing the beam path and the most essential optical components. (b) Measured photon flux at 118~nm and 200 kHz repetition rate for the SiC and Pt grating (blue circles and black squares respectively) as a function of Xe gas pressure in the tripling gas cell using a spot size of 175~$\mu$m (diameter). A maximum flux of $9\cdot 10^{12}$ photons/sec is achieved at a gas pressure of 8~mbar using the SiC grating.}%
\label{monofig}%
\end{figure}

The goal during the development of the present light source was to provide photons with sufficient energy so that the entire Brillouin zone of the high-temperature superconductors (high-$T_{C}$ materials) could be reached. However, keeping the photon energy below $\approx 11$ eV, which is the absorption threshold of LiF and MgF$_2$, would allow the monochromator to be separated from the ultra-high vacuum (UHV) system using windows, which in turn would dramatically simplify the design of the system. Therefore, it was decided that the target photon energy should be slightly below 11~eV. Achieving photon energies close to the 11~eV limit means that the HHG has to be performed in several steps which in turn requires a pulsed source with very good beam characteristics and very high peak power. The present light source is built around a picosecond pulsed laser (FUEGO from Time-Bandwidth Products AG). It provides 10~ps pulses at 1064 nm wavelength with a selectable repetition rate between 200~kHz and 8~MHz and a maximum average output power of 45~W at 200~kHz. Following the laser is a third-harmonic generation (THG) stage using second-harmonic generation (SHG) and sum-frequency generation (SFG) in two non-linear optical (NLO) crystals to convert the infrared light (1064~nm) to ultraviolet (UV) light of wavelength 354.7~nm. The efficiency of this tripling stage is 25-35\%, thus reducing the maximum average output to approximately 15~W at the lowest repetition rate. A second, consecutive frequency tripling stage based on THG in a xenon filled gas cell provides the final vacuum ultraviolet (VUV) light of wavelength 118.2~nm. In the gas cell, the THG is performed in a tight focusing geometry with a confocal parameter b=3.5~mm and a power density of $2\cdot 10^{12}$~W/cm$^{2}$ (at 200 kHz repetition rate) thus taking advandage of the negative-dispersion region of Xe just below the $5p$-$5d$ threshold at 119.2~nm \cite{Ward69}. The photons generated in this last THG process has an energy of 10.5~eV. With a work function of 4.5~eV, typical for many high-T$_{C}$ materials, a maximum kinetic energy of $\approx$6~eV can be reached, which corresponds to an in-plane wave vector of 1.25~\AA$^{-1}$ at 90 degrees emission angle. However, for practical reasons the maximum reachable emission angle is limited to approximately 75 degrees which sets the upper limit for the covered in-plane momentum to 1.21~\AA$^{-1}$.

Due to the strong absorption of VUV radiation in air, the gas cell is directly integrated in the monochromator housing which is kept under vacuum, see Fig.\ref{monofig}(a). The monochromator separates the 118.2~nm light from the remaining 354.7~nm after the last THG process and is based on a Littrow geometry consisting of a spherical re-focusing mirror followed by a plane blazed platinum coated grating or a plane laminar silicon-carbide (SiC) grating with line densities of 1200~lines/mm and 363~lines/mm, respectively. The grating is mounted in a motorized holder, enabling rotation around the horizontal and vertical axis respectively. This allows adjustment of the position of the beam in the measurement chamber, thus enabling optimum alignment of the beam relative the analyzer focus. Fig. \ref{monofig}(b) shows measured photon flux at 118~nm wavelength and 200~kHz repetition rate with a spot size of 175~$\mu$m as a function of Xe gas pressure for the platinum and SiC gratings. 
Since ultra-high vacuum (UHV) is required for most measurements of our interest, the vacuum of the monochromator is separated from the measurement chamber by a 500 $\mu$m LiF window which has a transmittance of 20-40 \% at 118.2~nm. This allows the pressure in the measurement chamber to be kept at $\sim~5\cdot 10^{-11}$~mbar even though the monochromator pressure can be as high as a few mbar. In order to avoid the problem of specular reflection of light into the ToF anlyzer, the geometry of the monochromator is such that the light exits at a 15 degree angle with respect to the horizontal plane. Consequently, the light is reflected above the entrance opening of the analyzer irrespective of the $\theta$-rotation of the sample. A rotatable $\lambda/2$ plate placed between the two tripling stages enables rotation of the polarization of the light between linear horizontal and linear vertical (or any desired value in between). The polarizer is motorized and can be set to change periodically between different suitable polarization directions during one measurement, thus minimizing the effect on photoemission intensity due to vanishing matrix elements \cite{Damascelli03,Kiss08}. Table \ref{lasertable} summarizes the main characteristics of the laser and monochromator.

\begin{table}
\begin{tabular*}{\columnwidth}{@{\extracolsep{\fill}} lr}
\multicolumn{2}{c}{Light source performance} \\ \hline
Wavelength & 118~nm\\
Photon energy	&	10.5~eV\\
Pulse length (1064~nm) & 10~ps\\
Spectral bandwidth (1064~nm)	& 0.255~nm\\
Repetition frequency & 200~kHz - 8~MHz\\
Energy resolution (118~nm) & $<$~1~meV\\
Spot size on sample (diameter) & $<$175~$\mu$m\\
Polarization & Linear (rotatable) \\
Maximum photon flux (SiC grating) & $9\cdot 10^{12}$  photons/sec\\
\hline \\
\end{tabular*}
\caption{\label{lasertable}Main characteristics of the light source used in the experimental setup. The value given for the maximum photon flux was measured using the SiC grating and 200~kHz repetition rate, for which the pulse energy is a maximum.}
\end{table}

\subsection{Vacuum system}
\begin{figure}%
\includegraphics[width=8.5cm]{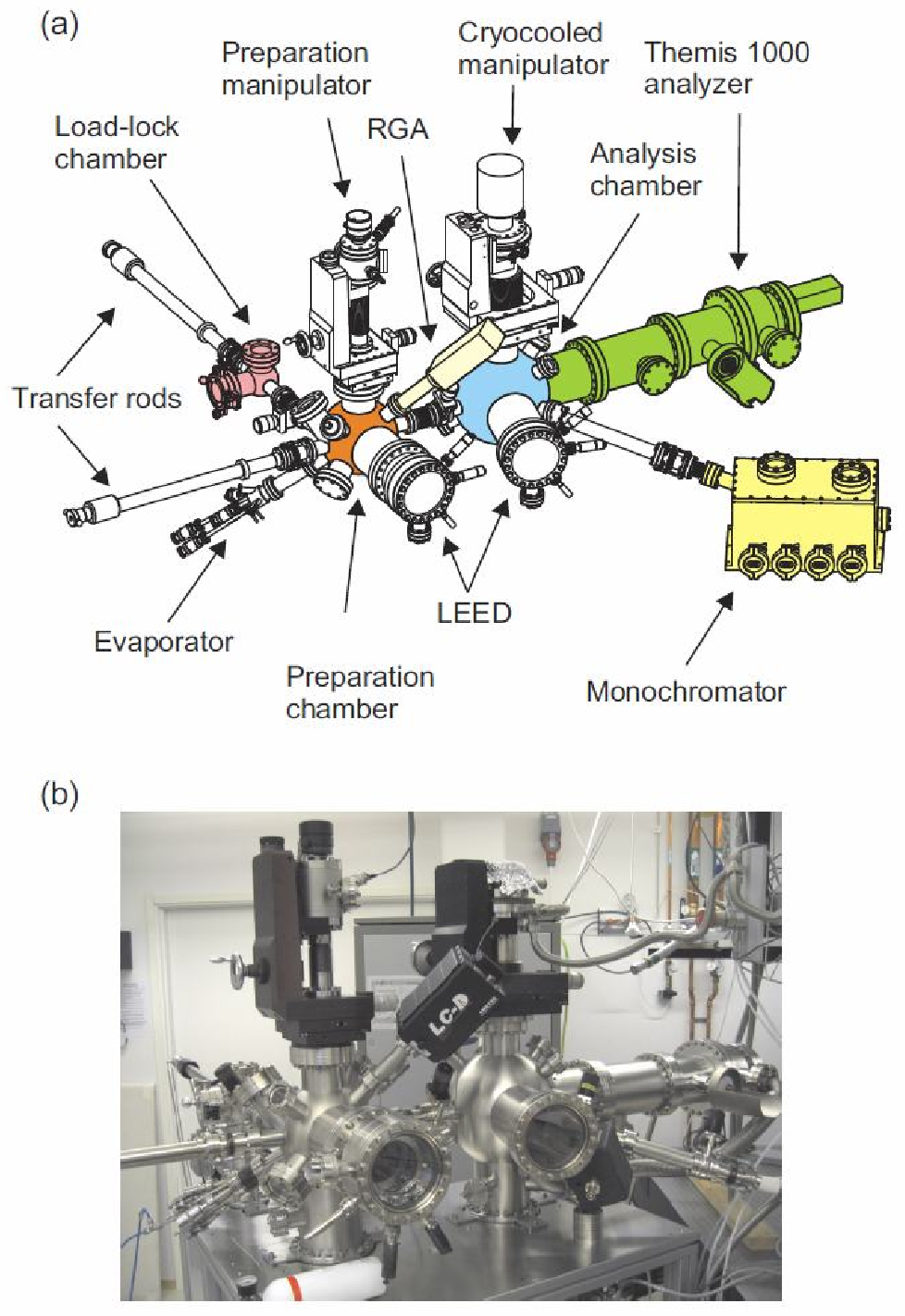}%
\caption{(Color online) Overview of the vacuum system. The most central parts of the system are the monochromator, the $\mu$-metal measurement chamber and the electron analyzer which in (a) are colored yellow, blue and green, respectively.}%
\label{syslayout}%
\end{figure}

The layout of the vacuum system can be seen in Fig. \ref{syslayout} and is typical for this type of system. It consists of three vacuum chambers; a) a load-lock for fast introduction of samples into the UHV system, b) a preparation chamber where standard pre- and post analysis and preparation of samples can be performed and c) the measurement (or analysis) chamber where the photoemission measurement takes place. The three chambers are separated by gate valves and samples are transferred between them by the use of linear transfer rods. 
The preparation chamber is equipped with an ion sputter gun for cleaning sample surfaces, a combined analysis unit for low energy electron diffraction (LEED) and Auger electron spectroscopy (AES), a three-target electron beam evaporator and a residual gas analyzer (RGA). A three-axis manipulator with $\theta$-rotation and a built in heating stage is used to move the samples between the sputtering, LEED and evaporation positions. The preparation chamber also contains a sample storage where up to four samples can be stored at once. Pumped by a turbomolecular pump, ion pump and a titanium sublimation pump (TSP) a base pressure of $8\cdot 10^{-11}$~mbar can be reached. 
Stainless steal is used as material for the load-lock and preparation chambers whereas the measurement chamber is made out of $\mu$-metal. The latter, in order to provide shielding of external magnetic fields which could disturb the flight paths of the photoemitted electrons. The measurement chamber is also equipped with a three-axis manipulator with $\theta$-rotation which is cooled by a closed-cycle cryocooler enabling sample temperatures in the range 7-350~K. A base pressure of $5\cdot 10^{-11}$~mbar is reached by pumping the chamber with a turbo, ion pump and TSP in combination with the cryo-pumping effect from the cooled manipulator.

\subsection{Electron analyzer}
\begin{figure*}%
\includegraphics[width=\textwidth]{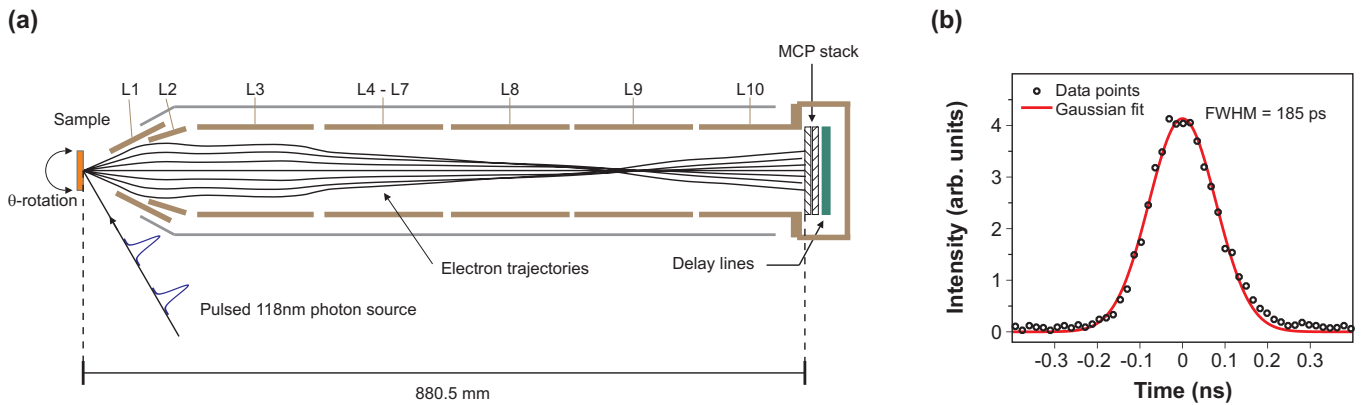}%
\caption{(Color online) (a) Schematic overview of the THEMIS 1000 electron analyzer with electron trajectories indicated by thin lines reaching from the sample surface to the delay-line detector (DLD) placed at the end of the electrostatic lens. The electrostatic lens elements are labeled L1-L10. When reaching the multi-channel plates (MCP) on the detector the electron signal is amplified and a pulse can be read out from the delay-lines. (b) Angle integrated spectrum in non-converted mode showing the peak of photoelectrons generated in the detector by light from the 10~ps laser pulse diffusely reflected off the sample. FWHM of this peak gives a measure of the time resolution of the system. In this case, the measured time-resolution was 185~ps.}%
\label{analyzerfig}%
\end{figure*}

One of the key components of the system is the electron spectrometer (THEMIS 1000 from SPECS GmbH) which consists of a cylindrical multi-element electrostatic lens and an electron detector. A schematic drawing of the analyzer including electron flight paths is shown in Fig.\ref{analyzerfig}(a). Electrons emitted from the sample surface within the acceptance angle of the analyzer are collected by the electrostatic lens and imaged onto a three-dimensional delay-line detector (DLD) placed at the end of the lens. Maximum acceptance angle is $\pm 15$~degrees and, since the entrance aperture of the analyzer is circular, electrons with any direction of the in-plane component of the momentum are collected. By reducing the acceptance angle of the lens the angular resolution can be increased. Therefore, the analyzer can be operated in four distinct modes with different acceptance angles; $\pm3$, $\pm4$, $\pm7$ and $\pm15$ degrees. 

The three-dimensional DLD (by Surface Concept GmbH) used to detect the electrons, has an active diameter of 40~mm. It consists of two Chevron mounted micro channel plates (MCP) which amplifies the incoming electrons and a crossed coil readout unit for determining their striking position (x,y) and time-of-impact (t) on the detector. A fraction of the photons in each laser pulse are diffusely reflected off the sample and hit the detector where they generate photoelectrons. These electrons give rise to a peak in the time histogram, see Fig.\ref{analyzerfig}(b). By measuring the full width at half maximum (FWHM) of this peak, the time-resolution of the system was found to be 185~ps. 
The DLD gives the position on the detector (x- and y-coordinate) and time-coordinate for each electron hit, the latter relative a time-zero defined by a trigger pulse from the laser. The time-of-flight principle uses the fact that the flight time ($t$) of one electron through the lens is given by $t^{2}=l^{2}m/2E_{kin}$, where $l$ is the distance the electrons travel along their trajectories from the sample, through the lens, to the detector, $m$ is the electron mass and $E_{kin}$ is the kinetic energy of the electron. Hence, electrons with high kinetic energy arrive first at the detector and the less energetic ones at later times $t$ giving a natural way of determining the energies of the electrons. Pre-calculated conversion matrices, generated by simulating the flight paths through the electrostatic lens for electrons emitted from the sample at different angles and energies, are used to convert the $x$, $y$ and $t$ coordinates of the actual detected electrons into emission angles and energies. Thus the spectrometer provides true three-dimensional detection, giving two spatial angles and energy at once. The DLD detector can operate at count rates up to 3~Mcps but is limited to one or two electrons per pulse. Also, in order to avoid space and image charge effects, the average number of photoelectrons generated per pulse should not exceed a few hundred \cite{Zhou05}. Taking into account the acceptance angle of the analyzer this implies that the count rate on the detector should be on the order of the repetition frequency of the laser, i.e. the count rate should not exceed the laser repetition rate.

In order to provide a field-free space between the sample and the analyzer a constant potential can be added to all lens elements, thus compensating for any work function difference between the sample and analyzer. Any electric field in this region would accelerate/retard the electrons and thereby influence their flight time. The main characteristics of the analyzer are presented in Table \ref{Themistable}.

\begin{table}
\begin{tabular*}{\columnwidth}{@{\extracolsep{\fill}} lr}
\multicolumn{2}{c}{Themis 1000 parameters} \\ \hline
Energy range & 0 - 3500 eV\\
Energy resolution & $<$ 500 $\mu$eV \\
Time resolution & 185~ps \\ 
Maximum angular acceptance & $\leq \pm 15^{\circ}$ \\
Maximum electron count rate on detector & $\leq$ 3 Mcps\\
\hline \\
\end{tabular*}
\caption{\label{Themistable}Key parameters for the THEMIS 1000 time-of-flight electron spectrometer. Note that the value given for the energy resolution is the theoretical resolution assuming a 100~ps time resolution and kinetic and pass energies of 5.9~eV and 1~eV, respectively, and thus differs from the experimentally determined energy resolution presented in this article.}
\end{table}

\section{Performance}

\subsection{Test of functionality}
\begin{figure}%
\includegraphics[width=8.5cm]{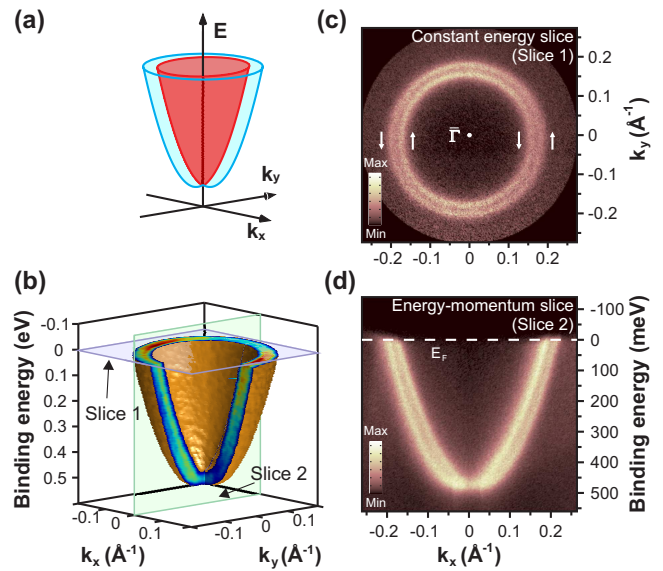}%
\caption{(Color online) Data showing the Au(111) surface state. (a) Principle drawing of the surface state consisting of two concentric parabolas with different spin directions (here represented by red and blue colors). (b) Raw data from the three-dimensional data matrix plotted as constant intensity surfaces (orange). Part of the data set (one quarter) is cut out in order to more clearly display the shape of the surfaces. One constant energy slice (Slice 1) and one energy-momentum slice (Slice2) are shown by horizontal (purple) and vertical (green) rectangles. (c) Constant energy slice corresponding to ``Slice 1'' in (b) showing the Fermi surface of Au(111). Spin directions for the two concentric circles are indicated by arrows (up/down). (d) Energy-momentum slice ($E$-$k_{x}$) at $k_{y}=0$ corresponding to ``Slice 2'' in (b) with the Fermi level given by the dashed horizontal line, labeled $E_{F}$.}%
\label{Audata}%
\end{figure}

To demonstrate the functionality of the system we have used the surface state of Au(111). This particular test sample was chosen because the parabolic surface state fits within the angular window covered by the wide-angle mode (WAM) of the analyzer ($\pm 15$ degrees acceptance angle). In addition, the surface state is split into two concentric parabolas, separated by $\Delta k_{\Vert}=0.023$~\AA$^{-1}$ \cite{LaShell} at the Fermi level, and thus tests the angle resolving capabilities of the spectrometer.
Previous photoemission experiments performed on Au(111) have observed the splitting of the parabolic free-electron like surface state \cite{LaShell,Reinert01,Nicolay01}. 
According to theoretical studies \cite{LaShell} the splitting is a result of the spin degeneracy being lifted by spin-orbit coupling, producing two bands with energy dispersion given by \cite{Petersen00}
\begin{eqnarray}
 E(k)=\frac{\hbar^2k^2}{2m}\pm \alpha k,
 \label{eq:spin}
\end{eqnarray}
where $\hbar$ is the Planck constant, $m$ the effective electron mass, $k$ the electron in-plane momentum and $\alpha$ the strength parameter which determines the size of the split. Effective electron masses for the two parabolas of $m=0.26~m_{e}$, where $m_{e}$ is the free electron mass, have been determined by previous experiments \cite{Reinert01}. The surface state electrons are confined in the z-direction, (i.e. direction perpendicular to the surface) and their spins are oriented in the surface plane ($k_{x}k_{y}$-plane), circulating around the $\bar{\Gamma}$-point in opposite directions for the two bands. Theory predicts these bands to be fully spin polarized, which is also supported by experiments \cite{Berntsen,Osterwalder}. 
Figure \ref{Audata} displays data of the Au(111) surface state acquired in one single measurement with the spectrometer described in this article. The single crystal used for the experiment was repeatedly cleaned by Ar-ion sputtering and annealed to $550^{\circ}$C until a sharp LEED pattern was observed. During measurement the sample temperature was 10 K and the base pressure $2\cdot 10^{-10}$mbar. The three-dimensional raw data matrix from a 15 minutes long measurement is visualized in Fig.\ref{Audata}(b) where constant intensity surfaces are drawn and a quarter of the data set is removed to more clearly display the shape of the surface state. From this data matrix two-dimensional slices in any direction can be made. Figure \ref{Audata}(c) and (d) show examples of typical constant energy (Slice 1 in Fig.\ref{Audata}(b)) and energy-momentum (Slice 2 in Fig.\ref{Audata}(b)) slices, respectively. In the former, showing the Fermi surface, two concentric circles separated by $\Delta k_{\Vert} = 0.021$~\AA$^{-1}$ are clearly visible, corresponding to the two bands with opposite spins. Arrows (up/down) in the figure indicate the spin direction. Fig.\ref{Audata}(d) shows the more familiar energy-momentum ($E$-$k$) slice, a spectrum similar to what one would get with a traditional hemispherical electron analyzer. As seen from this figure, the band dispersion follows the relation presented in Eq.(\ref{eq:spin}) with effective masses $m=0.28~m_{e}$ and the spin split is visible all the way from the Fermi level (labeled $E_{F}$ in the figure) towards the bottom of the parabolas, at a binding energy of $\approx$480~meV, until the split disappears at $k_{x}=0$. 
The functionality test presented here demonstrates the unique capability of this spectrometer, namely that a complete data set like the one displayed in Fig.\ref{Audata} can be collected in one single measurement with short acquisition time.

\subsection{Energy resolution}
\begin{figure}%
\includegraphics[width=8.5cm]{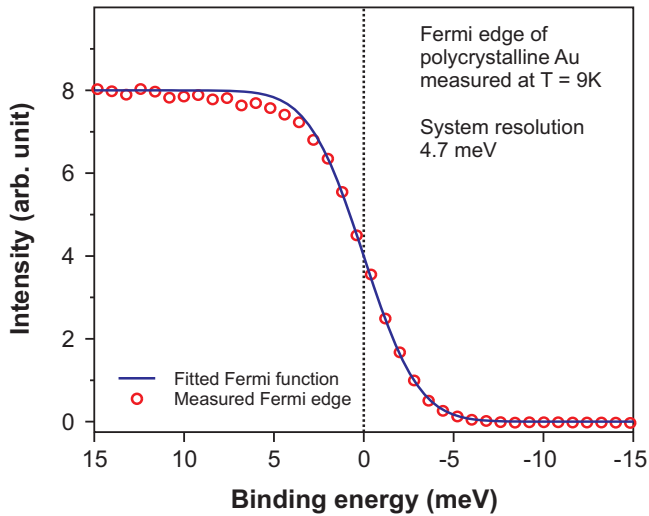}%
\caption{(Color online) The Fermi edge of polycrystalline Au measured at T=9~K. Circles (red) are measured data points. Solid (blue) line is a convolution of a Fermi function (at 9~K) and a Gaussian which then is fitted to the measured data. Obtained value for the FWHM=4.7~meV of the Gaussian is defined as the overall energy resolution of the system.}%
\label{energyres}%
\end{figure}

The overall energy resolution of the system contains contributions from both the light source and the analyzer. Thermal excitations in the sample will broaden the electron spectra, however at 9~K the temperature induced energy broadening is relatively small (3.1~meV). Since the laser has a spectral bandwidth of 0.255~nm at 1064~nm an energy broadening of 280~$\mu$eV is expected from the fundamental of the light source. Due to pulse chirping, the linewidth is not increased in the first THG step and consequently, at the final wavelength (118~nm) an energy broadening $<1$~meV is expected.
In order to determine the instrument energy resolution, i.e. the combined resolution of the light source and spectrometer, the Fermi edge of a polycrystalline Au sample has been measured. The sample was cleaned by repeated cycles of Argon ion sputtering and then cooled to 9~K for the measurement, which was performed under UHV conditions at a base pressure better than $3\cdot 10^{-10}$ mbar using the Low-Angular-Dispersion (LAD) mode of the analyzer ($\pm7$ degrees acceptance angle) and 1~eV pass energy. Figure \ref{energyres} shows the resulting data along with a fitted convolution between a Fermi distribution function of temperature 9~K and a Gaussian with FWHM~=~4.7~meV. This gives a total energy resolution of 4.7~meV for the system at a kinetic energy of 5.45~eV ($E_{F}$ in our measurement). 
Note that in order to produce the data points in Fig.\ref{energyres} we have integrated over all angles. The measured energy resolution is therefore the resolution from a data set integrated over the entire two-dimensional detector. Initial tests have shown that the energy resolution is sensitive to any misalignment of the light spot on the sample relative the optical axis and focus of the analyzer since this would make the Fermi surface appear tilted when it hits the detector. In such cases, integrating over the detector will result in a broader Fermi edge. Thus, by integrating over a smaller detector area a more precise value for the true energy resolution can be achieved.  Also, ground loops in the electric wiring of the system and electrical noise caused by the cryostat compressor is degrading the energy resolution of the analyzer. At present, work is done to minimize the influence of these sources of noise and therefore the energy resolution is expected to be further improved. 

\subsection{Angular resolution}
\begin{figure}%
\includegraphics[width=8.5cm]{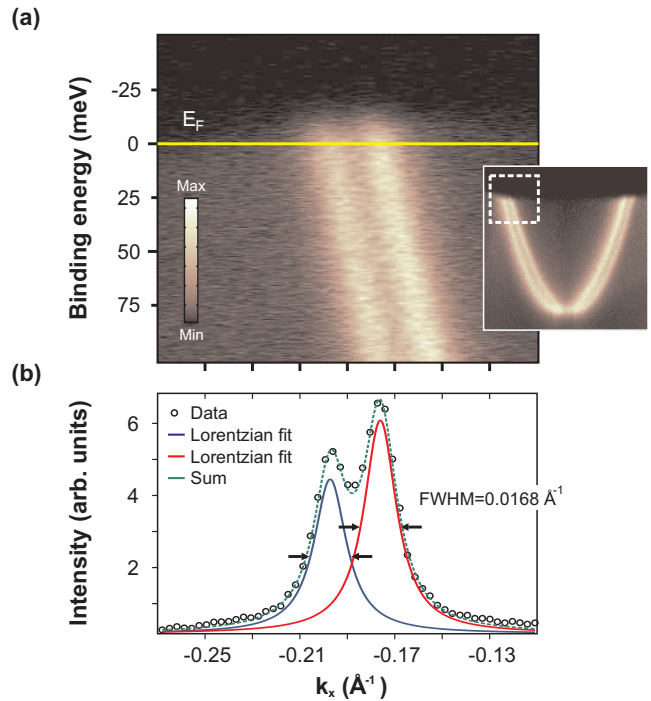}%
\caption{(Color online) (a) Zoomed view (see dashed rectangle in inset) of measured energy-momentum slice $E$-$k_{k}$ at $k_{y}=0$. Solid (yellow) line shows the Fermi level. (b) MDC at $E_{F}$ as indicated by solid line in (a). Circles (black) show measured data points, the two solid lines (red and blue) are fitted Lorentzian functions of equal widths, FWHM~=~0.0168~\AA$^{-1}$ (equals 0.8 degrees at $E_{k}$=5.55~eV). Dashed (green) curve is the sum of the Lorentzians. The fitted peak positions are separated by $\Delta k_{x}=0.021$~\AA$^{-1}$.}%
\label{angres}%
\end{figure}

In order to test the angular resolution of the system we have used the Au(111) data set discussed in the previous section and studied momentum distribution curves (MDCs) taken at the Fermi level. Figure \ref{angres}(a) shows a zoomed view of the area (given by dashed rectangle in inset) around the Fermi level in a $E$-$k_{x}$ slice through the data matrix at $k_{y}=0$. The solid (yellow) line in the figure indicates the Fermi level, labeled $E_{F}$, where the MDC studied in panel (b) has been taken. By fitting Lorentzians of equal width to the MDC, both peak positions and FWHM for the two peaks can be extracted. Obtained values for the peak position separation ($\Delta k_{E_{F}}$) and FWHM are 0.021~\AA$^{-1}$ and 0.0168~\AA$^{-1}$, respectively. The former is in good agreement with previous measurements \cite{LaShell,Reinert01,Nicolay01}, the latter corresponds to an angular spread of $0.8^{\circ}$ at a kinetic energy of 5.55~eV ($E_{F}$ in our measurements), also in agreement with previous measurements \cite{Nicolay01}. From this test we can not determine the exact angular resolution of the analyzer since the measured width of the peaks in the spectrum most likely is an inherent property of the sample and not necessarily limited by the instrument. Therefore, we conclude that the angular resolution for the $\pm 15^{\circ}$ mode (WAM-mode) at its worst is $0.8^{\circ}$, although the true angular resolution is expected to be less than this value. 

\subsection{The cuprate superconductor Bi2212}
\begin{figure}%
\includegraphics[width=8.5cm]{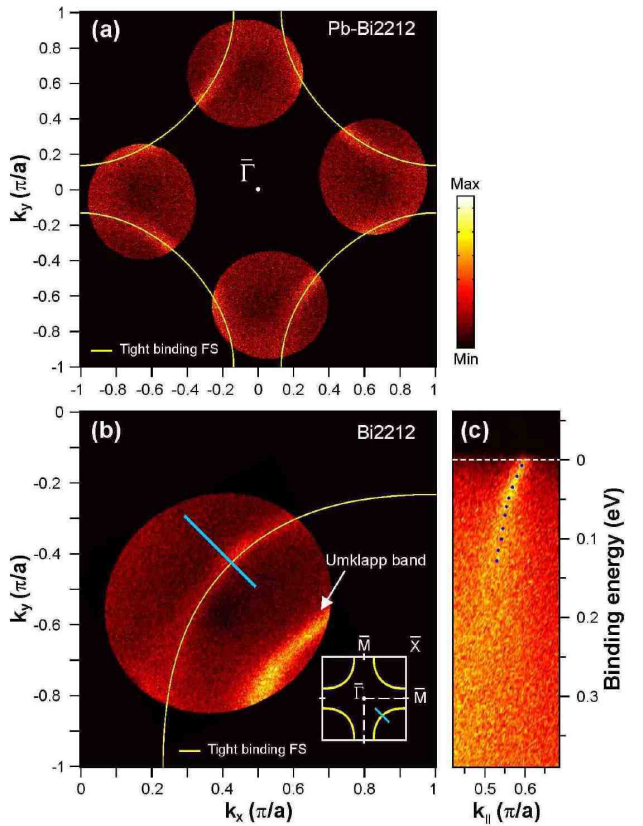}%
\caption{(Color online) (a) Part of the Fermi surface of Pb-Bi2212 measured at 9~K, data is symmetrized. Solid (yellow) line is a tight-binding fit to peak positions extracted from MDCs at the Fermi level. (b) A quarter of the Brillouin zone with part of the FS of Bi2212 measured at 9~K. Thin solid line is TB fit. The second band, running parallel to the main FS is the superstructure replica (Umklapp band) of the FS. Dashed square in inset shows where in the surface Brillouin zone the data has been taken. (c) $E$-$k_{\Vert}$ slice at position indicated by solid (blue) line in (b). Ten MDC peak positions down to $\sim~150$~meV below $E_{F}$ are marked with black dots.}%
\label{Bidata}%
\end{figure}
To test the system further, a more complex system, the high-temperature superconductor Bi$_{2}$Sr$_{2}$CaCu$_{2}$O$_{8+\delta}$ (Bi2212), is used. As part of the cuprate family this material has received a lot of attention and is one of the most intensively studied high-temperature superconductors. Key information for better understanding the superconductivity in this type of material is believed to be found in the energy interval around the Fermi level. Hence, Fermi surface (FS) mapping and studies of the superconducting gap along the FS are two examples of typical measurements that have been done over the years. The instrument described in this paper is ideal for these types of measurements since complete areas of momentum space can be mapped out in one single measurement. With a lattice constant a$\sim$3.8~\AA$^{-1}$ the antinode in Bi2212 at ($\pi$,0) is reached at $k_{\Vert}\sim$0.83~\AA$^{-1}$ and ($\pi$,$\pi$) at $k_{\Vert}\sim$1.17~\AA$^{-1}$ which falls within the $k$-space covered by the system with its 10.5~eV photon energy. Figure \ref{Bidata}(a) shows a plot of the Brillouin zone of Pb-Bi2212 with measured parts of the FS, integrated over 30~meV around $E_{F}$, together with a tight-binding (TB) calculation of the FS (solid thin line). Extracted peak positions from MDCs were used to fit the TB FS to the measured data. The circular shaped areas are data from one measurement which then is symmetrized in steps of $90^{\circ}$ to get a better overview of the shape of the full FS.
A similar type of measurement of Bi2212, covering the area around the node is displayed in Fig.\ref{Bidata}(b). This figure shows the 4:th quadrant of the Brillouin zone as indicated by the dashed square in the inset of Fig.\ref{Bidata}(b). Also in this case, a tight-binding FS (thin solid line) is fitted to peak positions of MDCs and displayed together with the data. In addition to the main FS there is also a second band visible which is the superstructure replica of the FS \cite{Ding96,Feng01}. A energy-momentum slice through the data matrix along the $\bar{\Gamma}$-$\bar{X}$ direction crossing the FS arch at the node, indicated by the solid line in Fig.\ref{Bidata}(b), is shown in Fig.\ref{Bidata}(c). Peak positions for a few MDCs down to approximately 150~meV below the Fermi level are marked in the figure from which the characteristic ''70-meV kink'' \cite{Bogdanov00} can be observed. Both the Pb-Bi2212 and the Bi2212 samples used for the measurement were single crystals which were glued directly onto a Cu sample holder and cleaved at a temperature of 9~K under ultra-high vacuum conditions (pressure $\sim 1\cdot 10^{-10}$~mbar). The measurement was performed using the $\pm 15$ degrees wide-angle mode of the analyzer.

\section{Summary}
\begin{table}
\begin{tabular*}{\columnwidth}{@{\extracolsep{\fill}} lr}
\multicolumn{2}{c}{System performance} \\ \hline
Photon energy & 10.5 eV \\
Maximum reachable $k_{\Vert}$	&	1.21~\AA$^{-1}$ \\
Overall energy resolution & $<$~4.7 meV \\
Angular resolution & $<$~0.8 degrees \\
$k$-space resolution (at $E_{kin}=5.55$~eV)  &$<$~0.02 {\AA}$^{-1}$ \\
Temperature range & 7-350 K \\
Spot size (circular) &  0.024 mm$^2$ \\
Maximum photon flux (at 118~nm) & $9\cdot 10^{12}$ photons/sec \\
\hline \\
\end{tabular*}
\caption{\label{performance}Summary of the performance of the laser-based angle-resolving time-of-flight (LARTOF) photoemission setup.}
\end{table}

Over the years, going from one-dimensional to two-dimensional photoelectron analyzers has dramatically improved the use of photoemission as a technique for studying complex materials and their electron structure. Together with recent advances in laser based photon sources this transition has made very high resolution studies of complex materials feasible. The new instrument presented here takes the development one step further, adding one more angular dimension to the measurements, resulting in three-dimensional photoemission spectra while at the same time pushing the energy limit of a laser-based, high repetition rate, photon source for ARPES to 10.5~eV. The result is a system capable of reaching all parts of the Brillouin zone in many materials and providing high resolution maps of the energy dispersion over a two-dimensional area in $k$-space in one single measurement. Consequently, momentum-momentum band dispersion can be studied directly with this setup. The data presented in this article shows examples of typical measurements which can be performed with this instrument and verifies that the system has energy and angular resolutions, see Table \ref{performance}, which enables measurements of materials of strong current interest like, for example, cuprate high-temperature superconductors.

\section{Acknowledgements}
The instrumental development described here was made possible through a grant from the K\&A Wallenberg foundation as well as financial support from the Swedish Research Council (VR). The authors would like to thank A. \"{O}stlin for his work on the SiC grating and also Dr. T. U. Kampen and T. Kunze at SPECS GmbH for their technical assistance during the commissioning of the instrument. We also acknowledge the group of M. Golden at the Van der Waals Zeeman Institute, University of Amsterdam for providing the Bi2212 and Pb-Bi2212 samples.

\bibliography{lartof}

\end{document}